\documentclass[aps, prd, preprint, nofootinbib, amsfonts, floatfix]{revtex4}

\usepackage{graphicx}
\usepackage{amsmath,amsfonts}
\usepackage{amssymb}
\usepackage{dsfont}
\usepackage{url}
\usepackage{braket}
\usepackage{slashed}
\usepackage{epsfig,color}
\usepackage[dvipsnames]{xcolor}
\newcommand{\be}{\begin{eqnarray}}
\newcommand{\ee}{\end{eqnarray}}

\newcommand\nn{\nonumber}
\newcommand{\Tr}{{\rm Tr}}
\newcommand{\Str}{{\rm Str}\,}
\newcommand{\mat}{\left ( \begin{array}{cc}}
\newcommand{\emat}{\end{array} \right )}
\newcommand{\vect}{\left ( \begin{array}{c}}
\newcommand{\evect}{\end{array} \right )}
\newcommand{\tr}{{\rm Tr}}

\newcommand{\Sdet}{{\rm Sdet}\,}
\newcommand{\PR}{{\Psi^{(R)}}}
\newcommand{\PL}{{\Psi^{(L)}}}

\definecolor{red}{rgb}{1.00, 0.00, 0.00}

\definecolor{blue}{rgb}{0.00, 0.00, 1.00}
\definecolor{green}{rgb}{0.20, 0.6, 0.1}

\definecolor{darkgreen}{rgb}{0.0, 0.4, 0.0}

\let\Re\relax
\DeclareMathOperator{\Re}{Re}
\let\Im\relax
\DeclareMathOperator{\Im}{Im}

\long \def \blockcomment #1\endcomment{}

\begin{document}

\title{New term in effective field theory at fixed topology}

\author{M.~Kieburg${}^1$, M.~Lauritzen,${}^2$ B.T.~S{\o}gaard${}^2$, K. Splittorff${}^2$}
\affiliation{${}^1$School of Mathematics and Statistics, University of Melbourne,\\
813 Swanston Street, Parkville, Melbourne VIC 3010, Australia \\ ${}^2$The Niels Bohr Institute, University of Copenhagen, Blegdamsvej 17, DK-2100, Copenhagen {\O}, Denmark 
}

\date{\today}
\begin{abstract}
A random matrix model for lattice QCD which takes into account the positive definite nature of the Wilson term is introduced. The corresponding effective theory for fixed index of the Wilson Dirac operator is derived to next to leading order. It  reveals a new term proportional to the topological index of the Wilson Dirac operator and the lattice spacing. The new term appears naturally in a fixed index spurion analysis. The spurion approach reveals that the term is the first in a new family of such terms and that equivalent terms are relevant for  the effective theory of continuum QCD.
\end{abstract}

\maketitle


\section{Introduction}

The duality between random matrix theory (RMT) and low energy effective field theory (EFT) has revealed a plethora of insights in physical systems as diverse as quantum chromo dynamics (QCD) \cite{ShuryakVerbaarschot,3foldway} and topological solid state systems which realise Majorana fermions \cite{Beenakker}. 
Most results obtained address average spectral properties of a central operator for the system in question, such as the Hamiltonian or the Dirac operator, but also universal parametric correlations can be obtained from RMT and EFT \cite{GHS,Szafer-Altshuler,BR,Mehlig,AFK,ADSV}. 
The duality between the two approaches is highly valuable as some questions may be technically easier to address in one of the two frameworks. In addition some questions only have a natural formulation in one approach. One example of this is the effect of the Hermiticity-properties of the operator in question: In the RMT formulation the Hermiticity properties are obvious since the operator is directly present, on the contrary in the EFT approach these properties are hidden in the low energy constants (LEC) \cite{KSV}. 
In this work we investigate how the positive definite nature of an operator explicitly appearing in RMT affects the dual EFT. Remarkably this will allow us to extend the duality and show that random matrix theory can be used to discover new terms in the low energy effective theory. The new terms found are intimately linked to the topological properties of the theory and appear in the effective action for fixed topology.  Studies of effective actions with fixed topology are of great value \cite{LS} and may for example be used to determine the LEC's of the effective theory \cite{DHSS}.

The physical realization we study here is lattice regularised QCD. In particular, we focus on the Wilson term \cite{Wilson} which is essential in order to remove the Fermonic doublers from lattice QCD, see e.g.~\cite{GattringerLang}. The Wilson term is a covariant Laplacian \cite{Wilson} and thus positive definite \cite{Adams}, see appendix \ref{latticeApp}. The explicit symmetry breaking of the Wilson term is well understood in EFT \cite{SharpeSingleton,RS,BRS}, however, the effect of the positive definite nature of the Wilson term on the EFT is studied here for the first time. We will introduce a new random matrix model (RMM)\footnote{We use the abbreviation RMT when referring to the general topic of random matrix theory and RMM when referring to a specific random matrix model.}, which takes into account the fact that the Wilson term is positive definite. A general method to derive the next to leading order terms in the EFT from the RMM is then developed and used. The resulting EFT uncovers a new term in the effective action for fixed topology. The term which is linear in the lattice spacing and the topological index is similar to an axial mass term. We show that the new term appears naturally from a fixed $\nu$ spurion analysis and that it is the first in a family of such new terms.

We use the EFT to explain why an order $a$-improvement of lattice actions does not only move the Dirac eigenvalues closer to the origin but at the same time also decreases the width of the distributions, as was observed in lattice QCD simulations \cite{DHS}. 

Finally we consider continuum QCD. We use the fixed $\nu$ spurion approach to show that new terms also appear in the effective action for continuum QCD at fixed topology. We show that the new terms are fully consistent with the effective theory at fixed $\theta$-angle \cite{Gasser:1983yg}, despite that there are no new terms in this action.

Longer derivations are given in the appendices along with a  discussion of the axial anomaly at non-zero lattice spacing.
 
\section{The new matrix model}

The new RMM with a positive definite analogue of the Wilson term is defined as (the parameter $a>0$ is the analogue of the lattice spacing)
\be
\label{rmtmodel}
&Z^\nu&=\int d Wd AdB \, P(A,B,W) \, \det (D_W +M) , \\
&& D_W   =  \begin{bmatrix}
a A&W\\
-W^\dagger& a B
\end{bmatrix}, \quad M=\begin{bmatrix}
M_R & 0\\
0 & M_L
\end{bmatrix}. \nn
\ee
 The matrix $D_W$ is the RMT analogue of the Wilson Dirac operator and the diagonal term proportional to $a$ corresponds to the Wilson term.  In order to take into account the positive definite nature of the Wilson term the diagonal terms are positive definite Hermitian matrices $A\in{\rm Herm}_+(N_R)$ and $B\in{\rm Herm}_+(N_L)$. We are using a chiral basis and the integers $N_R$ and $N_L$ indicate the number of right and left handed states. As in the original chiral random matrix theory \cite{ShuryakVerbaarschot} the matrix $W$ is complex, $W\in \mathbb{C}^{N_R \times N_L}$. The matrix structure ensures that the partition function has fixed index $\nu$ of the Dirac operator 
\be
\nu \equiv N_L-N_R = \sum_{j} \langle\psi_j |\gamma_5| \psi_j \rangle ,
\ee
where $| \psi_j \rangle$ are the eigenvectors of $D_W$.

The matrices are chosen to be distributed along 
\be
P(A,B,W)\propto {\det}^{\nu_A} A\,{\det}^{\nu_B} B\,  e^{-(N_L+N_R)/2 \, (\tr W^\dagger W+\tr A+\tr B)} \ .
\label{P(ABW)}
\ee
This weight is not constrained by chiral symmetry and the Gaussian form is chosen for simplicity. As the spurion argument in section \ref{sec:spurion} below shows the new term in the corresponding EFT is determined by the chiral symmetries of the $D_W$ rather than by the weight. The positive definite matrices  $A$ and $B$ model the Wilson term and the exponents $\nu_A,\nu_B\geq0$ allow for a variation of the level repulsion in the spectra of $A$ and $B$\footnote{Equivalently for integer $\nu_A$ (or $\nu_B$) we could have modelled the level repulsion by writing $A$ (or $B$) as $A=X^\dagger X$ with $X\in\mathbb{C}^{(N_R+\nu_A)\times N_R}$ distributed as $P(X)\propto e^{-(N_L+N_R)/2\Tr X^\dagger X}$, which exactly corresponds to $A$ distributed as $P(A)\propto {\det}^{\nu_A} A\, e^{-(N_L+N_R)/2\Tr A}$ \cite{goodman}. Thus by writing the repulsion as determinants we generalize to $\nu_A$ and $\nu_B$ real and positive. We would like to emphasise that the dyadic structure $X^\dagger X$ mimics the structure of the Wilson term, see appendix \ref{latticeApp}}. These exponents are not constrained by symmetry and will combine into two LEC's in the EFT.  Finally, the fermion masses are $M_R,M_L\in\mathbb{C}^{N_f\times N_f}$.

The RMM introduced above reduces to the original RMM for continuum QCD \cite{ShuryakVerbaarschot} in the limit $a=0$. In this limit $A$ and $B$ drops out of the Wilson Dirac operator and the partition function factors into an integral over $W$, which equals the partition function of the original $a=0$ RMM, and  an integral over $A$ and $B$ which is equal to $1$. Note that contrary to the RMM for $a\neq0$ introduced in \cite{DSV} it is not possible to absorb the sign of $a$ into the random matrix it is multiplied by, and as we show below this introduces odd terms in $a$ in the effective theory.  

This RMM is invariant under parity which interchanges $N_L\leftrightarrow N_R$ (i.e.~$\nu\rightarrow -\nu$) and $\nu_A\leftrightarrow\nu_B$. Note that though parity changes the sizes of $W$, $A$ and $B$  the overall size
\be
N \equiv N_L+N_R  \ ,
\ee
of $D_W$ is fixed. Additionally  $\nu_A-\nu_B$ is odd under this transformation while $\nu_A+\nu_B$ is even. We introduce $w_t$ and $w_M$ such that $\nu_A-\nu_B=w_t\nu+\nu$ and $\nu_A+\nu_B=w_M\geq0$, and as we show below $w_t$ and $w_M$ are the natural combinations which become LEC's in the dual EFT.

\section{The effective theory at fixed $\nu$}

The dual EFT is obtained from the RMM in two steps (for a detailed derivation see appendix \ref{app:RMTtoEFT}). The first step is exact. We express the determinants as an integral over Fermionic variables $\Psi^{(R/L)}$ and then average over $A$, $B$ and $W$. After using the superbosonization formula~\cite{superbos} to exchange the dyadic matrices $\Psi^{(R/L)\dagger}\Psi^{(R/L)}$ with the unitary matrices $U_{(R/L)}$ we obtain 
\be
Z^\nu & \sim & \int_{[\text{U}(N_f)]^2}d\mu (U_R)d\mu (U_L) {\det}^{-N_R}U_R{\det}^{-N_L}U_L \nn \\
&&\times{\det}^{\nu_A+N_R}({\mathds 1}_{N_f}+aU_R){\det}^{\nu_B+N_L}({\mathds 1}_{N_f}+aU_L) \label{exact}
\\ &&\times
 \exp\left[\frac{N_R+N_L}{2}\left(\tr U_RU_L+\tr(M_RU_R+M_L U_L)\right)\right]  \ ,\nn
\ee
where the integration is over the normalised Haar measure. In the second step we define the counting $a\sim m \sim 1/\sqrt{n}$,  where we introduced $n$ with $N_R=n$ and $N_L=n+\nu$  to simplify the notation. We then substitute $U_R=\sqrt{U_a}U$ and $U_L=U^{-1}\sqrt{U_a}$, expand the massive modes like $U_a=\exp(iH/\sqrt{n})={\mathds 1}+iH/\sqrt{n}-H^2/(2n)+\ldots$ and keep all terms up to order $1/\sqrt{n}$. Finally, we integrate over the Hermitian matrix $H$ and get the EFT up to order $1/\sqrt{n}$, 
\be
\begin{split}\label{Znu}
 Z^\nu =& \int_{{\rm U}(N_f)}\hspace*{-0.cm}d\mu(U) {\det}^\nu U   e^{(N/2+N_f/8)\tr(UM_R^{(a)}+M_L^{(a)} U^{-1})-(Na^2/4)\tr(U^2+U^{-2})-(N/16)\tr\left(UM_R^{(a)}+M_L^{(a)} U^{-1}\right)^2}\\
	&\times e^{(N/192)\tr\left(UM_R^{(a)}+M_L^{(a)} U^{-1}\right)^3+(Na^2/8)\tr\left(UM_R^{(a)}+M_L^{(a)} U^{-1}\right)(U^2+U^{-2})}\\
	&\times e^{(Na^3/6)\tr(U^3+U^{-3})+(w_t\nu a /2)\tr(U-U^{-1})} .
\end{split}
\ee
Note that the shifted mass matrix $M_{R/L}^{(a)}=M_{R/L}+a+w_m a/N$ results naturally from the positivity of the Wilson term.  

While the Gaussian form of the weight (\ref{P(ABW)}) makes this direct computation possible, we stress the Gaussian form is not essential for the terms in the EFT. Likewise, the values of the exponents $\nu_A$ and $\nu_B$ are not important for the terms generated in \eqref{Znu}, even if the determinantal factors are omitted $\nu_A=\nu_B=0$, we obtain the same terms in the EFT (with $w_t=-1$). A general $\nu_A$ and $\nu_B$, however, generates the free LEC $w_t$. For a detailed discussion of this universality in the context of RMT at $a=0$ see \cite{ADMN}. 

A word on the counting before we proceed: We here make use of the $p$-regime counting for $m$ and derivative terms therefore also enters at the leading orders in the EFT. Since it is zero dimensional the RMM does not generate dynamical terms. However, the virtue of the RMT approach used here is that we can explicitly derive the corresponding EFT from the RMM and in this way obtain terms in the EFT that may have been overlooked in the standard approach to EFT.

\section{The new term in the EFT and Spurion analysis at fixed $\nu$}
\label{sec:spurion}

All terms apart from the last in the effective action of (\ref{Znu}) also appear in \cite{SharpeSingleton,RS,BRS} and this allows us to identify $N$ as the dimensionless volume where the dimension is set by a LEC's multiplying each term in the effective action.  
However, the last term in~\eqref{Znu}
\be
w_t\nu \frac{a}{2}\tr(U-U^{-1})
\label{new-term}
\ee
has not appeared previously. Note that this new term takes the form of an axial mass proportional to $\nu a$.


Let us try to understand why the new term has not appeared in effective actions previously.
From the RMT side the term could not be generated by the model of~\cite{DSV} since this by construction was even in $a$. 
From the EFT side~\cite{SharpeSingleton,RS,BRS}, one writes down the most general effective action consistent with the symmetries order by order in $a$ of the  Symanzik action at fixed $\theta$-angle. However, one discards all terms that are total derivatives~\cite{Luscher:1996sc,Luscher:1998pe}. Here, we consider the theory in a sector with fixed index of the Dirac operator, therefore in the associated continuum expansion the topological density, which is a total derivative, will integrate to $\nu$. Including the total derivatives associated with the fixed topology opens for possible new terms in the Symanzik expansion, which in turn gives rise to new terms in the effective theory. As we now show the new term, $w_t \nu a\tr(U-U^{-1})/2$, has just the right structure from a spurion perspective.  


On this end we now extend the standard spurion approach to fixed $\nu$ and show that the new term, (\ref{new-term}), arises naturally in this fixed $\nu$ spurion analysis.  The basic rules for a fixed $\nu$ spurion analysis are the same as used for the fixed $\theta=0$ spurion analysis in \cite{SharpeSingleton,RS,BRS}: First, since the explicit breaking of chiral symmetry by the Wilson term at leading order in $a$ can be restored provided that we spurion transform $a\to g_R a g_L^\dagger$, the lattice spacing can enter the effective action only through invariant combinations such as $\tr(aU^\dagger)$ and $\tr(a^*U)$. Second, the effective action must be invariant under parity which interchanges $U\leftrightarrow U^\dagger$. In addition  in the fixed $\nu$ spurion analysis we must take into account that volume splits into a parity even part $N_L+N_R$ and a parity odd part $\nu=N_L-N_R$. Hence we naturally have the invariant combinations $(N_L+N_R)\tr(a^*U+aU^\dagger)$ and $\nu\tr(a^*U-aU^\dagger)$ to leading order in $a$.
From a spurion perspective at fixed $\nu$ the new term found is therefore as natural as the ordinary linear term in $a$. Of course since typically $N_L+N_R\gg |N_L-N_R|$ the new term is of higher order.

\section{The effective theory at fixed $\theta$}

In order to check that the EFT with the new term is physically consistent with the standard EFT at fixed $\theta$, we now derive the partition function at fixed vacuum angle $\theta$.  
From the spurion approach no new terms are expected in $Z_\theta$, since for fixed $\theta$ the volume does not split naturally in a parity even and odd part. As we now show by explicitly deriving $Z_\theta$ this is indeed the case.

We use the relation 
\be\label{defZth}
Z_\theta(m,a) \equiv {\sum}_{\nu=-\infty}^\infty e^{i\nu\theta}Z^\nu(m,a) \ ,
\ee
to define the partition function at fixed $\theta$ for non-zero $a$. Note that despite the new term proportional to $\nu$ in the EFT we still have $Z_\nu=Z_{-\nu}$, 
and thus $Z_\theta=Z_{-\theta}$. 
For notational simplicity we set $N_f=1$ where $U=e^{i\tilde\theta}$ and set the mass matrix to $m_a\equiv M_R^{(a)}=M_L^{(a)}=m+a+w_m a/N\in\mathbb{R}$, meaning without axial mass source term. The sum over $\nu$ in~\eqref{defZth} imposes the constraint $\theta+\tilde\theta+w_t a\sin(\tilde\theta)=0$ which when expanded in $a$ yields
\be
 \tilde{\theta}=-\theta+w_t a\sin(\theta)-w_t^2a^2\sin(2\theta)/2+\mathcal{O}(a^3) . 
\ee
We thus obtain for the partition function up to $\mathcal{O}(1/\sqrt{n})$
\be\label{Zth-Nf1}
 &Z_\theta&= \exp\left[(Nm_a+m_a/4+w_t a)\cos(\theta)-Na^2/2\cos(2\theta)+w_t Nm_a a\sin^2(\theta)-N m_a^2/4\cos^2(\theta)\right] \nn \\
&&\times \exp\left[Nm_a^3/24\cos^3(\theta)-w_t N m_a^2a/2\cos(\theta)\sin^2(\theta)\right] \\
	&&\times \exp\left[Na^2m_a/2(\cos(\theta)\cos(2\theta)+w_t\sin(\theta)\sin(2\theta)-w_t^2\cos(\theta)\sin^2(\theta))\right] \nn \\
&&\times \exp\left[Na^3/3(\cos(3\theta)-3w_t\sin(2\theta)\sin(\theta))\right] . \nn
\ee

Despite the new term at fixed $\nu$, the partition function at fixed $\theta$ has no new terms. $Z_\theta$ is, as it should be, perfectly consistent with the standard spurion approach where the volume does not split into a parity even and odd part. 

The effective theory with fixed $\theta$ just obtained allows us to compute the topological susceptibility 
\begin{equation}
\langle\nu^2\rangle  =  -\partial_\theta^2\log( Z_\theta)|_{\theta=0} =Nm_a+\mathcal{O}(1).
\end{equation}
Therefore, we  have $\langle\nu^2\rangle\sim \sqrt{N}$ in the counting considered, and hence the new term is in fact typically enhanced by a factor $N^{1/4}$.

\section{Two equivalent formulations}

It is of course possible also to go back to the partition function $Z^\nu$ we started from using \cite{LS}
\be\label{Znu-from-Ztheta}
Z^\nu(m) & = & \frac{1}{2\pi} \int_{-\pi}^\pi d\theta \, e^{-i\nu\theta}Z_\theta(m) \ .
\ee
If we insert $Z_\theta$ from  (\ref{Zth-Nf1}) we have a formulation of $Z^\nu$ which contains only standard terms
\be\label{Znu-alt}
Z^\nu(m) & = & \frac{1}{2\pi} \int_{-\pi}^\pi d\theta\, e^{-i\nu\theta}\exp\left[-a\cos(\theta)\right]\nn\\
	&&\times\exp\left[N(m+a)\left(\cos(\theta)+a\sin^2(\theta)-\frac{3}{2}a^2\cos(\theta)\sin^2(\theta)\right)\right]\\
	&&\times\exp\left[-\frac{Na^2}{2}\left(\cos(2\theta)+2a\sin(\theta)\sin(2\theta)\right)\right]\nn\\
	&&\times\exp\left[-\frac{N}{4}(m+a)^2\left(\cos^2(\theta)+2a\cos(\theta)\sin^2(\theta)\right)+\frac{N}{24}(m+a)^3\cos^3(\theta)\right]\nn\\
	&&\times\exp\left[\frac{Na^2}{2}(m+a)\cos(\theta)\cos(2\theta)+\frac{Na^3}{3}\cos(3\theta)\right]\nn\\
		&&\times\exp\left[\frac{a}{4}\cos(\theta)\right] \ .\nn		
\ee
However, when we shift the integration variable 
\be\label{shift-th}
\theta=\theta' + a\sin(\theta') \ ,
\ee
and expand in $a$ we recover the expression (\ref{Znu}) for $Z^\nu$ we started from.
Note that the Jacobian 
\be
\frac{\partial \theta}{\partial\theta'} = 1 +  a\cos(\theta')\sim \exp\left[a\cos(\theta')\right] \ ,
\ee
cancels the factor from the $\delta$-function. 
The conclusion is remarkable: $Z^\nu$ has two formulations at the given order, one including only standard terms and one including the new term.

The two partition functions (\ref{Znu}) and (\ref{Znu-alt}) are equal at the given order. This is similar to asymptotic expansions which may have different and yet equivalent expressions to a given order \cite{Murray}. The different formulations may be useful for example when determining the low energy constants of the EFT.

\section{Application of the EFT}

The EFT (\ref{Znu}) may be used to derive properties of the real eigenvalues $\lambda_k$ of the original Wilson Dirac operator. For example, we may get the distribution of the chiralities over the real eigenvalues of the Wilson Dirac operator~\cite{ADSV} 
\be\label{rhochi-def}
\rho_{\chi}(\lambda)  \equiv \left\langle \sum_k\delta(\lambda-\lambda_k){\rm sign}\langle{\psi_k}|\gamma_5 |\psi_k\rangle\right\rangle ,
\ee
through the relation~\cite{ADSV}
\be\label{rho-from-chi}
\rho_{\chi}(\lambda) 
& = & \frac{1}{\pi}{\rm Im}\left[\Sigma(m=-\lambda)\right]  ,
\ee
where the Green function of $D_W$ is
\be
\Sigma(m)\equiv\left\langle \tr\frac{1}{D_W+m}\right\rangle \ . 
\ee
Hence we compute $\rho_{\chi}$ from the EFT by the supersymmetry technique, see \cite{ADSV} for details on this method. 
For simplicity, we consider the quenched case employing the relation
\be
\Sigma(m)= \lim_{m'\to m}\partial_m\log Z_{1|1}(m|m'),
\ee
where the quenched partition  function, $Z_{1|1}$, comprises a single valence fermion and boson and is given by a supersymmetric integral. We now show that the overall structure of the calculation motivates the counting and allows us to understand the effect of order $a$ improvement on the real modes of $D_W$.  The details of the computations are described in appendix \ref{app:SUSY}.

The counting  $a\sim m \sim 1/\sqrt{n}$ employed here is relevant as the positive definite order $a$ term will move the real eigenvalues of the Wilson Dirac operator from the origin to values of order $a$. Hence, the Fermion mass $m$, which in the quenched supersymmetric partition function becomes the eigenvalue, cf.~(\ref{rho-from-chi}), must be of the same order. Otherwise the resulting eigenvalue density will only probe regions where no eigenvalues appear. Order $a$ improvement of the lattice action will primarily reduce the LEC of the order $a$ term in the EFT action \cite{Golterman-rev} and thus correspondingly the magnitude of the eigenvalue. If the order $a$ improvement is so accurate that the LEC of the order $a$ term in the EFT becomes of order $1/\sqrt{n}$, then the relevant counting for the eigenvalues and hence quark mass in the quenched supersymmetric partition function becomes $m\sim 1/n$. Thus accurate order $a$ improvement will connect back to the standard $\epsilon$-counting where $m\sim 1/n$ and $a\sim1/\sqrt{n}$. The EFT derived here, in this way, allow us to monitor the effect of the order $a$ improvement. 

\begin{figure}[t!]
\begin{center}
\includegraphics{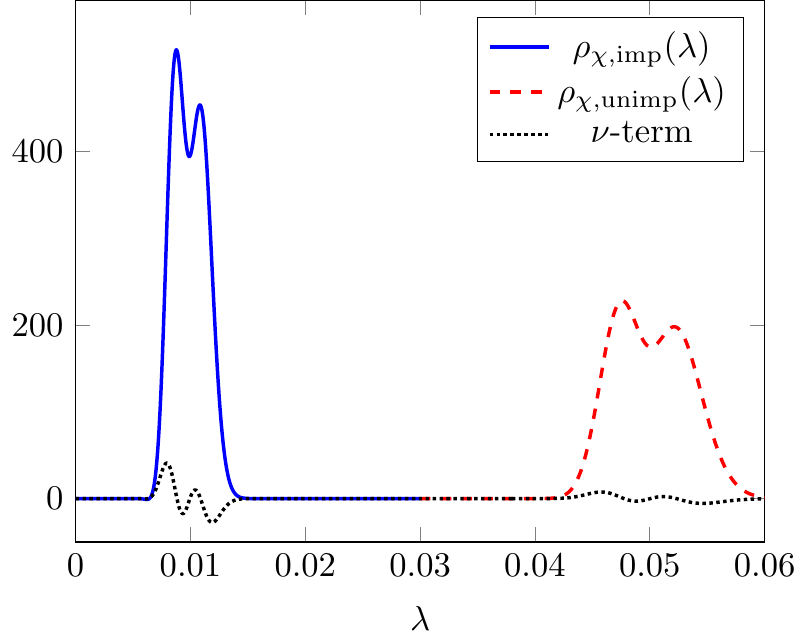}
\caption{The distribution of the chiralities over the real eigenvalues of the Wilson Dirac operator, $\rho_{\chi}(\lambda)$ defined in Eq.~(\ref{rhochi-def}).  As the LEC of the leading order $a$-term is reduced (as for order $a$-improvement), the distribution not only moves toward the origin it also becomes more narrow. Compare to the lattice QCD data of figure 1 in the second entry of \cite{DHS}. The dotted curves display the effect of the new term obtained by subtracting from the full result the result without the new term. The parameters are $\nu=2$, $n=100$, $a=1/\sqrt{n}$, $w_t=1$ and $w_m=0$. For the unimproved curve the LEC of the leading order term in $a$ is enhanced by a factor 5.}
\label{fig:rhochi}
\end{center}
\end{figure}

A simple example of this is as follows: Order $a$-improvement of the Wilson Dirac operator naturally moves the eigenvalues of the Wilson Dirac operator closer to the origin, since the order $a$-term acts as a mass. Surprisingly, however, as observed in \cite{DHS} the distribution of the real eigenvalues also becomes narrower when the Wilson Dirac operator is order $a$-improved. The action of (\ref{Znu}) offers a natural explanation: The width of the distribution of the real modes comes from the order $1$ and $1/\sqrt{n}$ terms in the action which, we note,  includes $m$. As the action is order $a$ improved the relevant eigenvalue and hence the relevant $m$ used in the supersymmetric method decreases. This in turn suppress  the order  $1$ and $1/\sqrt{n}$ terms resulting in a narrower (and hence more continuum like) distribution of the real modes of the Wilson Dirac operator, see Fig.~\ref{fig:rhochi} where the effect of the new term is also explicitly shown. 

An other example of how the EFT can be applied is given in appendix \ref{app:anomaly}, where we show that it can be used to analyse the spectral contributions to the axial anomaly. 

\section{The effective theory of continuum QCD}

It is natural to ask if the new term, (\ref{new-term}), is special to the effective theory for lattice QCD or whether it is also relevant in the effective theory for continuum QCD \cite{Gasser:1983yg} at fixed $\nu$ \cite{LS}. To answer this let us consider the fixed $\nu$ spurion analysis in the continuum. Since $\nu$ is fixed the volume again splits into a parity even and a parity odd part, and thus it is as natural to have a new term $\nu\tr(m^*U-mU^\dagger)$ in the effective action as it is to have the usual mass term $(N_L+N_R)\tr(m^*U+mU^\dagger)$.   

We now show how the new term $\nu\tr(m^*U-mU^\dagger)$ appears in the effective action at fixed $\nu$ even if we start from an effective theory at fixed $\theta$ which only includes standard terms. To simplify we neglect all terms except the mass and the $L_7$ term of \cite{Gasser:1983yg}. We make use the relation $Z_{\theta}(m)=Z_{\theta=0}(me^{i\theta/N_f})$ \cite{LS} and start from
\be\label{ZEFTtheta_a=0}
Z_{\theta}(m) &\sim & \int_{{\rm SU}(N_f)} d\mu(\tilde{U}) 
 \\
&&\times
	\exp\left[\frac{1}{2}V\Sigma m\tr(e^{-i\theta/N_f}\tilde{U}+e^{i\theta/N_f}\tilde{U}^\dagger)-4Vm^2B_0^2L_7\tr^2(e^{-i\theta/N_f}\tilde{U}-e^{i\theta/N_f}\tilde{U}^\dagger)\right] \ , \nn
\ee
with $B_0=\frac{\Sigma}{F_\pi^2}$ \cite{Gasser:1983yg}. We now integrate as in (\ref{Znu-from-Ztheta}) 
\be\label{ZEFTnu_a=0}
Z^\nu(m) &\sim &  \frac{1}{2\pi} \int_{-\pi}^\pi d\theta \, e^{-i\nu\theta}\int_{{\rm SU}(N_f)} d\mu(\tilde{U})  \\
&&\times
	\exp\left[\frac{1}{2}V\Sigma m\tr(e^{-i\theta/N_f}\tilde{U}+e^{i\theta/N_f}\tilde{U}^\dagger)
-4Vm^2B_0^2L_7\tr^2(e^{-i\theta/N_f}\tilde{U}-e^{i\theta/N_f}\tilde{U}^\dagger)\right] \ . \nn
\ee
In this form $Z^\nu$ is expressed using standard terms. However, we can shift the integration variable
\be\label{theta-x_a=0}
\theta=\theta' + i\frac{x}{2}\tr(e^{-i\theta'/N_f}\tilde{U}-e^{i\theta'/N_f}\tilde{U}^\dagger) \ ,
\ee
and choose $x= 16mN_fB_0^2L_7/\Sigma\sim 1/\sqrt{V}$, such that the contribution from the standard mass term is canceled. With this change of variables we obtain 
\be\label{ZEFTnu_a=0-shifted}
&Z^\nu(m) &\sim   \frac{1}{2\pi} \int_{-\pi}^\pi d\theta' \, e^{-i\nu\theta'}\int_{{\rm SU}(N_f)} d\mu(\tilde{U}) \,  \\
	&&\times \exp\left[\frac{8mB_0^2L_7}{\Sigma}\tr(e^{-i\theta'/N_f}\tilde{U}+e^{i\theta'/N_f}\tilde{U}^\dagger)+\frac{1}{2}V\Sigma m\tr(e^{-i\theta'/N_f}\tilde{U}+e^{i\theta'/N_f}\tilde{U}^\dagger)\right] \nn \\
	&&\times \exp\left[\nu m\frac{16N_fB_0^2L_7}{\Sigma}\tr(e^{-i\theta'/N_f}\tilde{U}-e^{i\theta'/N_f}\tilde{U}^\dagger)\right] \ . \nn
\ee
(The first term in the second line is the Jacobian.) After the integration over $\theta'$ and $\tilde U$ the partition function in the standard representation, \eqref{ZEFTnu_a=0}, is equal to the partition function with the new term, \eqref{ZEFTnu_a=0-shifted}, to the order we work at. Again the different formulations are similar to how asymptotic series may have different but equivalent expressions \cite{Murray}.

The LEC of the new term is proportional to $L_7$ which according to best fits \cite{Bijnens} is non-zero. We therefore conclude that new term is relevant for the effective action for continuum QCD at fixed topology. The new equivalent formulation of the EFT can perhaps be useful in determining the low energy constant $L_7$. In particular we stress that in the new formulation (\ref{ZEFTnu_a=0-shifted}) the squared trace in (\ref{ZEFTnu_a=0}) is changed for terms linear in the Goldstone field.

\begin{figure}[t]
\begin{center}
\includegraphics{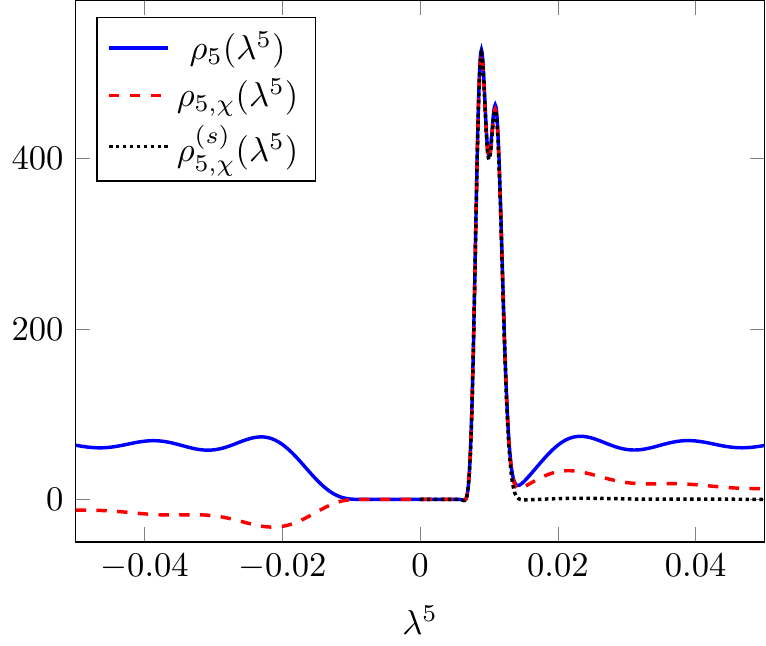}
\caption{The eigenvalue density $\rho_5(\lambda^5)$ (blue solid), the distribution of the chiralities $\rho_{5,\chi}(\lambda^5)$ (red dashed) and the symmetrized version $\rho_{5,\chi}^{\rm (s)}(\lambda^5)$  (black dotted), for $m=0$, $a=0.1/\sqrt{n}$, $n=100$, $w_t = 1$, $w_m = 0$ and $\nu=2$. Note that $\rho_{5,\chi}^{\rm (s)}(\lambda^5)$ is fully dominated by the near topological modes. \label{fig:rho5-chi}}
\label{default}
\end{center}
\end{figure}

\section{Generalisations}

The fixed $\nu$ spurion analysis allows us to identify a family of new terms in the effective action at fixed topology. All we need to ensure is that the action must respect parity and that volumes $N_L+N_R$ and $N_L-N_R$ must only enter to first power such that the action is extensive. For example $\nu\tr(m^*U-mU^\dagger)\tr(m^*U+mU^\dagger)$ is a perfectly valid higher order term in the continuum effective action at fixed topology.

Related we may ask if the shift of variable as in (\ref{shift-th}) and (\ref{theta-x_a=0})  is unique or if there are other possibilities. On this end we note that the shift of variable (\ref{theta-x_a=0}) is Hermitian, respects the standard spurion structure and is parity odd. Other shifts are possible and as long as they respect this all new terms generated automatically conserve the fixed $\nu$ spurion rules.

Finally we note that effective actions for QCD where topology is directly linked to the Goldstone field has a long history, see eg.~\cite{DiVecchiaVeneziano}, but we stress that the terms considered here are of a different nature.

\section{Summary}

Our analysis of a new RMM for lattice QCD with a positive definite analogue of the Wilson term has revealed a new term in effective actions at fixed topology. We have extended spurion analysis to fixed topology and used this to show that the term is the first in a new family of terms. While the new terms where discovered in the context of the EFT at non-zero lattice spacing we have shown that they are relevant even for the effective theory of continuum QCD. In particular, we have explicitly shown how the new terms at fixed topology can arise starting from an effective action at fixed $\theta$ including only standard terms. The effective theory including the new term has been used to discuss the effect of order $a$ improvement as well as to obtain new insights in the spectral contributions to the axial anomaly. Both of these insights have been obtained by explicitly deriving the relevant spectral correlation functions using the supersymmetric technique. 

It would be most interesting to explore if the formulation of  the effective theory with the new term present can be used to obtain better bounds on the physical parameters which appear as low energy constants in the effective theory. 
\bigskip

\acknowledgements
 We thank J.~Greensite, J.~Verbaarschot and P.H.~Damgaard for discussions ass well as J.~Bijnens for correspondence. BTS acknowledges the hospitality while visiting the University of Melbourne.

\appendix

\section{Properties of the Wilson Dirac operator in lattice QCD}
\label{latticeApp}
For completeness this appendix reviews some of the properties of the Wilson Dirac operator following the setup described in \cite{Adams}. In 4 dimensional lattice QCD we discretize the Euclidean spacetime with some lattice spacing $a$ such that we consider a lattice $\mathcal T\subset \mathbb{R}^4$. The spinor fields $\psi(x)$ live on the lattice sites and take values in $\mathbb C^4\otimes\mathbb C^{N_c}$, where $N_c$ is the number of colors in the SU($N_c$) gauge symmetry. We can further equip this space of spinor function with an inner product
\begin{align}
	\braket{\psi_1|\psi_2} \equiv a^4 \sum_{x\in \mathcal T} \psi_1(x)^\dagger \psi_2(x).\label{innerprodpos}
\end{align}
The gauge field resides on the bonds of the lattice as parallel transporters $U_\mu(x)$ taking values in the gauge group. In particular $U_\mu(x)$ parallel transports a spinor from $x+a e_\mu$ to $x$, where $e_\mu$ is a unit vector in the $\mu$th direction. This allows us to write covariant forward (backward) difference operators
\begin{align}
	\nabla^+_\mu \psi(x) = U_\mu(x)\psi(x+ae_\mu)-\psi(x),\qquad \nabla^-_\mu\psi(x)=\psi(x)-U_\mu(x-ae_\mu)^\dagger\psi(x-ae_\mu).
\end{align}
We can then compute the Hermitian conjugate of $\nabla^+_\mu$ with respect to the inner product \eqref{innerprodpos}
\begin{align}
	\braket{\psi_1|\nabla^+_\mu \psi_2} &=a^4\sum_{x\in\mathcal T} \psi_1(x)^\dagger(U_\mu(x)\psi_2(x+ae_\mu)-\psi_2(x)) \\
	&=a^4\sum_{x\in\mathcal T} \left[(U_\mu(x)^\dagger\psi_1(x))^\dagger\psi_2(x+ae_\mu)-\psi_1(x)^\dagger\psi_2(x)\right]\nonumber\\
	&=a^4\sum_{x\in\mathcal T}(U_\mu(x-ae_\mu)^\dagger\psi_1(x-ae_\mu)-\psi_1(x))^\dagger\psi_2(x)=\braket{-\nabla^-_\mu\psi_1|\psi_2},\nonumber
\end{align}
where we shifted the sum in the last line. A similar computation holds for $\nabla^-_\mu$ and hence we have $(\nabla^\pm_\mu)^\dagger=-\nabla^\mp_\mu$. This implies that $\nabla_\mu = \frac{1}{2}(\nabla_\mu^-+\nabla_\mu^+)$ is anti-hermitian $(\nabla_\mu)^\dagger = -\nabla_\mu$ which further renders the Dirac operator $\slashed{\nabla} = \gamma^\mu\nabla_\mu$ anti-hermitian (in Eucledian signature $\gamma^\mu$ are hermitian). We can define the covariant ``Laplacian'' $\Delta$ that occurs in the Wilson term
\begin{align}
	\Delta \equiv \sum_\mu (\nabla^+_\mu)^\dagger\nabla_\mu^+ ,
\end{align}
which is clearly Hermitian and semi positive-definite as
\begin{align}
	\braket{\psi|\Delta \psi} = \sum_\mu \braket{\nabla^+_\mu\psi|\nabla^+_\mu\psi}\ge 0.
\end{align}
Finally, we can state the full Wilson Dirac operator
\begin{align}
	D_W = \gamma^\mu \left(\frac{1}{a}\nabla_\mu\right)+\frac{a}{2}\left(\frac{1}{a^2}\Delta\right).
\end{align}

\section{From RMM to EFT}
\label{app:RMTtoEFT}

In this appendix we give details on how to derive the EFT from the RMM. As above we consider the random matrix model defined in Eq.~(\ref{rmtmodel})
and first rewrite the determinant as fermionic Gauss integral
\begin{align}
	Z^\nu \propto &\int dWdAdBd\Psi\,{\det}^{\nu_A}A\,{\det}^{\nu_B}B\exp\left[-(n+\nu/2)(\tr WW^\dagger+\tr A+\tr B)\right]\\
	&\times\exp\left[\tr (M_R\PR^\dagger\PR+M_L\PL^\dagger\PL)-a\tr A\PR\PR^\dagger-a\tr B \PL\PL^\dagger\right]\nonumber\\
	&\times\exp\left[-\tr W\PL\PR^\dagger+\tr W\PR\PL^\dagger\right]\nonumber.
\end{align}
Where $\Psi^{(R/L)}$ has indices in both flavor and ``RMT'' space i.e. $\Psi^{(R)}_{if}$ with $i=1,\ldots,n$ and $f=1,\ldots,N_f$. Further we write $\PL\PR^\dagger$ as shorthand for $\Psi^{(L)}_{if}\Psi^{(R)*}_{jf}$ and $\PL^\dagger\PR$ for $\Psi^{(L)*}_{if}\Psi^{(R)}_{ig}$. One can now show through  series expansion that $\det(\mathds{1}_n+\PR\PR^\dagger)={\det}^{-1}(\mathds{1}_{N_f}+\PR^\dagger\PR)$, where the inverse is due to the anti-commutative nature of the $\Psi$'s. We can then solve the integrals over $A$ and $B$ using $\int_{\text{Herm}_+(n)}dA\,{\det}^\nu A\, e^{-\tr AC} \propto {\det}^{-\nu-n}C$ and the $W$-integral using standard Gaussian integrals
\begin{align}
Z^\nu \propto &\int d\Psi\,{\det}^{-\nu_A-n}[(n+\nu/2)\mathds{1}+a\PR^\dagger\PR]{\det}^{-\nu_B-n-\nu}[(n+\nu/2)\mathds{1}+a\PL^\dagger\PL]\\
&\times\exp\left[-\frac{1}{n+\nu/2}\tr\PL\PR^\dagger\PR\PL^\dagger+\tr (M_R\PR^\dagger\PR+M_L\PL^\dagger\PL)\right]\nonumber.
\end{align}
Using the superbosonization formula we exchange ${\Psi^{(R/L)}}^\dagger\Psi^{(R/L)}\rightarrow (n+\nu/2)U_{R/L}\in (n+\nu/2)\text{U}(N_f)$ and simultaneously factor out $(n+\nu/2)$ of the determinants as
\begin{align}
Z^\nu \propto & \int_{\text{U}(N_f)^{\otimes 2}}d\mu(U_R)d\mu(U_L)\,{\det}^{-n}U_R{\det}^{-n-\nu}U_L {\det}^{\nu_A+n}(\mathds{1}+aU_R){\det}^{\nu_B+n+\nu}(\mathds{1}+aU_L)\\
&\times \exp\left[\left(n+\frac{\nu}{2}\right)(\tr U_RU_L+\tr(M_RU_R+M_LU_L))\right]\nonumber.
\end{align}
We can then shift the integral as $U_L\rightarrow U_R^{-1}U_L$ followed $U_R\rightarrow \sqrt{U_L}U_R$ and identify $U=U_R$ and $U_a=U_L$
\begin{align}
	Z^\nu \propto & \int_{\text{U}(N_f)^{\otimes 2}}d\mu(U)d\mu(U_a)\,{\det}^{\nu}U{\det}^{-n-\nu/2}U_a {\det}^{\nu_A+n}(\mathds{1}+a\sqrt{U_a}U){\det}^{\nu_B+	n+\nu}(\mathds{1}+aU^{-1}\sqrt{U_a})\nonumber\\
	&\times \exp\left[\left(n+\frac{\nu}{2}\right)(\tr U_a+\tr(M_R\sqrt{U_a}U+M_LU^{-1}\sqrt{U_a}))\right].
\end{align}
After that a saddlepoint approximation can be performed around $U_a=\mathds{1}$, where we use the  counting $M_R\sim M_L\sim a \sim \frac{1}{\sqrt{n}}$. Thus we expand as $U_a=\mathds{1}+\frac{i}{\sqrt{n}}H-\frac{1}{2n}H^2+\ldots$ with $H$ Hermitian and collect terms up to $\frac{1}{\sqrt{n}}$
\begin{align}
	Z^\nu \approx & \int d\mu(U)dH\,{\det}^{\nu}U\\
	&\times\exp\left[\frac{N}{2}\tr(M_R^{(a)}U+U^{-1}M_L^{(a)})-\frac{na^2}{2}\tr(U^2+U^{-2})\right] \nn \\
	&\times\exp\left[\frac{na^3}{3}\tr(U^3+U^{-3})+w_t\nu \frac{a}{2}\tr(U-U^{-1})\right]\nonumber\\
	&\times\exp\left[-\frac{1}{2}\tr H^2+\frac{i\sqrt{n}}{2}\tr(H(M_R^{(a)}U+U^{-1}M_L^{(a)}))\right]\nonumber\\
	&\times\exp\left[-\frac{i}{6\sqrt{n}}\tr H^3 -\frac{1}{8}\tr(H^2(M_R^{(a)}U+U^{-1}M_L^{(a)}))-\frac{i\sqrt{n}a^2}{2}\tr(H(U^2+U^{-2}))\right]\nonumber,
\end{align}
where we define $\nu_A-\nu_B = w_t\nu+\nu$, $\nu_a+\nu_b=w_m$ and the shifted mass matrix $M^{(a)}_{R/L} = M_{R/L}+a+w_Ma/N$. We can then shift the integral in $H$ by the linear term (it is sufficient to shift by the linear term in the third line) and expand the $\frac{1}{\sqrt{n}}$-suppressed terms
\begin{align}
Z^\nu \propto & \int d\mu(U)dH\,{\det}^{\nu}U\\
&\times\exp\left[\frac{N}{2}\tr(M_R^{(a)}U+U^{-1}M_L^{(a)})-\frac{na^2}{2}\tr(U^2+U^{-2})\right] \nn \\
	&\times\exp\left[\frac{na^3}{3}\tr(U^3+U^{-3})+\frac{w_t\nu a}{2}\tr(U-U^{-1})\right]\nonumber\\
&\times\exp\left[-\frac{1}{2}\tr H^2-\frac{n}{8}\tr((M_R^{(a)}U+U^{-1}M_L^{(a)})^2)\right]\nonumber\\
&\times\left[1-\frac{n}{48}\tr((M_R^{(a)}U+U^{-1}M_L^{(a)})^3) -\frac{1}{8}\tr(H^2(M_R^{(a)}U+U^{-1}M_L^{(a)}))\right.\nonumber\\
&\qquad\left.+\frac{n}{32}\tr((M_R^{(a)}U+U^{-1}M_L^{(a)})^3)+\frac{na^2}{4}\tr((M_R^{(a)}U+U^{-1}M_L^{(a)})(U^2+U^{-2}))\right],\nonumber
\end{align}
where we neglected odd power of $H$ as they vanish due to symmetry (after the shift). The average over $H$ is essentially a constant $\braket{1}=1$ and a non-trivial term $\braket{\tr H^2 A}=N_f\tr A$ so that we get after re-exponentiation
\begin{align}
Z^\nu \propto & \int d\mu(U)\,{\det}^{\nu}U\exp\left[\left(\frac{N}{2}+\frac{N_f}{8}\right)\tr(M_R^{(a)}U+U^{-1}M_L^{(a)})-\frac{Na^2}{4}\tr(U^2+U^{-2})\right]\\
&\times\exp\left[-\frac{N}{16}\tr(M_R^{(a)}U+U^{-1}M_L^{(a)})^2+\frac{N}{192}\tr(M_R^{(a)}U+U^{-1}M_L^{(a)})^3\right]\nonumber\\
&\times\exp\left[\frac{Na^2}{8}\tr(M_R^{(a)}U+U^{-1}M_L^{(a)})(U^2+U^{-2})+\frac{Na^3}{6}\tr(U^3+U^{-3})+\frac{w_t\nu a}{2}\tr(U-U^{-1})\right]\nonumber.
\end{align}

\section{The supersymmetric technique}
\label{app:SUSY}

In order to obtain the various spectral correlation functions $\rho_\chi$, $\rho_5$ etc. from a given effective theory, we employ the so called supersymmetric (SUSY) technique, see eg.~\cite{DOTV,efetov}. The actual partition function of interest has $N_f$ flavors. The idea of SUSY is to add two valance flavors, one of fermionic statistics and one of bosonic with masses and axial masses $m$, $z$ and $m'$, $z'$ respectively. The partition function including the new valence flavors takes the form
\begin{align}
Z^\nu_{N_f+1|1}&= \int D[A_\mu]_\nu\,e^{-S_{YM}[A_\mu]}\frac{\det(D_W+m+\gamma^5 z)}{\det(D_W+m'+\gamma^5z')}\prod_{f=1}^{N_f}\det(D_W+m_f),
\label{partition-susy}
\end{align}
where the path integral is taken over the sector of topological index $\nu$. The main observation is that in the limit $m'\rightarrow m$ and $z'\rightarrow z$ the partition function \eqref{partition-susy} coincides with the original without valence flavors. Thus we can derive certain statistics with respect to the original ensemble by differentiating $Z^\nu_{N_f+1|1}$ with respect to $m$ and $z$ and subsequently taking the aforementioned limit. In particular we can calculate the following resolvents
\begin{align}
	\Sigma(m,z)&=\lim_{\substack{m'\rightarrow m\\z'\rightarrow z}} \frac{\partial}{\partial m}\log Z^\nu_{N_f+1|1}=\Braket{\tr\frac{1}{D_W+m+\gamma^5z}},\\
	\Sigma_5(m,z)&=\lim_{\substack{m'\rightarrow m\\z'\rightarrow z}} \frac{\partial}{\partial z}\log Z^\nu_{N_f+1|1}=\Braket{\tr\frac{\gamma^5}{D_W+m+\gamma^5z}}=\Braket{\tr\frac{1}{D_5+z}}.
\end{align}
From the resolvents we can directly obtain the spectral correlation functions. As an example we work out the relation for $\rho_5(\lambda^5)$ i.e. the spectral density of $D_5=\gamma^5(D_W+m)$. As $D_5$ is Hermitian we can write the expression for $\Sigma_5$ in the eigenbasis of $D_5$, letting $z=-\lambda^5-i\epsilon$ we have
\begin{align}
	\Sigma_5(m,z=-\lambda^5-i\epsilon) = \Braket{\sum_{k}\frac{1}{\lambda^5_k-\lambda^5-i\epsilon}}= i\pi \Braket{\sum_{k}\delta(\lambda^5_k-\lambda^5)}+\ldots,
\end{align}
where we use the Sokhotski–-Plemelj theorem and '$\ldots$' denotes the real part. This immediately gives us the relation
\begin{align}
	\rho_5(\lambda^5) \equiv \Braket{\sum_{k}\delta(\lambda^5-\lambda^5_k)} = \frac{1}{\pi}\Im\Sigma_5(m,z=-\lambda^5+i\epsilon).\label{rho5}
\end{align}
Similarly for $D_5$ one can derive the following relation
\begin{align}
	\rho_{5,\chi}(\lambda)\equiv \left\langle\sum_{k}\delta(\lambda^5-\lambda^5_k)\braket{\psi^5_k|\gamma^5|\psi^5_k}\right\rangle = \frac{1}{\pi}\Im\Sigma(m,z=-\lambda^5+i\epsilon).\label{rho5chi}
\end{align}
For the Wilson operator we can also compute the distribution of the real eigenvalues of $D_W$ weighted by either the sign or the value of the chirality yielding $\rho_\chi(\lambda)$ \cite{ADSV} and $\rho_{\text{cont},\chi}(\lambda)$
\begin{align}
	\rho_\chi(\lambda) &\equiv \left\langle\sum_{k:\lambda_k\in\mathbb{R}}\delta(\lambda-\lambda_k)\operatorname{sign}\braket{\psi_k|\gamma^5|\psi_k}\right\rangle = \frac{1}{\pi}\Im\Sigma(m=-\lambda,z=-i\epsilon),\\
	\rho_{\text{cont},\chi}(\lambda) &\equiv \left\langle\sum_{k:\lambda_k\in\mathbb{R}}\delta(\lambda-\lambda_k)\braket{\psi_k|\gamma^5|\psi_k}\right\rangle = \frac{1}{\pi}\Re\Sigma(m=-i\lambda-\epsilon,z=0).
\end{align}
In particular we can use the above to derive spectral densities of chiral effective theories, where the partition function is expressed as a integral over the Goldstone manifold ${\rm U}(N_f)$. 

To accommodate for the valance flavors we extend the integration manifold to $Gl(N_f+1|1)$ \cite{DOTV}. The Lagrangian of (\ref{Znu}) is effectively promoted to being supersymmetric by replacing traces and determinants with their SUSY counterparts i.e. $\tr\rightarrow \Str$ and $\det\rightarrow\Sdet$. This leaves the question of parameterization of $U\in {\rm Gl}(N_f+1|1)$ and its accompanying measure. For simplicity we consider the quenched case $N_f=0$ and use the parameterization of \cite{DOTV}
\begin{align}
	U = \begin{bmatrix}
		e^{i\theta}&0\\
		0& e^s
	\end{bmatrix}\exp\begin{bmatrix}
		0&\alpha\\
		\beta&0
	\end{bmatrix},\qquad d\mu(U) = \frac{d\theta}{2\pi}\,ds\,d\beta\, d\alpha ,
\end{align}
with $\theta \in [-\pi,\pi]$, $s\in(-\infty,\infty)$ and $\alpha$, $\beta$ Grassmannian. Hence we arrive at the SUSY integral representation of the the graded partition function
\begin{align}
Z^\nu_{1|1} = & \int \frac{d\theta}{2\pi}\,ds\,d\beta\,\,{\Sdet}^{\nu}U\exp\left[i\frac{N}{2}\Str(M_R^{(a)}U-U^{-1}M_L^{(a)})-\frac{Na^2}{4}\Str(U^2+U^{-2})\right]\nonumber\\
&\times\exp\left[-\frac{N}{16}\Str(M_R^{(a)}U-U^{-1}M_L^{(a)})^2-\frac{iN}{192}\Str(M_R^{(a)}U-U^{-1}M_L^{(a)})^3\right]\\
&\times\exp\left[-\frac{iNa^2}{8}\Str(M_R^{(a)}U-U^{-1}M_L^{(a)})(U^2+U^{-2})-\frac{iNa^3}{6}\Str(U^3-U^{-3})\right]\nonumber\\
&\times\exp\left[\frac{iw_t\nu a}{2}\Str(U+U^{-1})\right]\nonumber,
\end{align}
where we shifted the integral as $U\rightarrow iU$ to ensure convergence \cite{DSV,ADSV}. At this point it is straightforward, but rather tedious to expand out the Grassmann part of the exponent and solve the Grassmann integral. This leaves the partition function as an integral over the $\theta$ and $s$, which can be differentiated to find $\Sigma$ and $\Sigma_5$. The remaining two-fold integral can be numerically integrated to yield the spectral functions $\rho_{5}$, $\rho_{5,\chi}$, $\rho_\chi$ and $\rho_{\text{cont},\chi}$ as described above.

\section{The axial anomaly}
\label{app:anomaly}

In this appendix we use the EFT to discuss the spectral contributions to the axial anomaly.  As shown by Fujikawa~\cite{fujikawa} the response of the Fermionic measure to an axial transformation
\be
|\psi\rangle \to |\psi'\rangle=e^{i\gamma_5 \alpha}|\psi\rangle, \qquad 
\langle \psi| \to \langle\psi' |= \langle\psi| e^{i\gamma_5 \alpha} \nn
\ee
includes a nontrivial Jacobian $J$. 
Here, we extend the argument of Fujikawa to non-zero lattice spacing by using the eigenvalues and eigenvectors 
\be
D_5|\psi^5_k\rangle = \lambda^5_k|\psi^5_k\rangle \ ,
\ee
of the Hermitian Wilson Dirac operator, $D_5 = \gamma_5(D_W+m)$.
We follow the proof of Fujikawa and obtain
\be
J & = &  \exp\left(-i \alpha\sum_k\langle{\psi^5_k}|\gamma_5 |\psi^5_k\rangle\right) \ .
\ee
Fujikawa regulates the infinite sum as
\be
J & = &  \lim_{\Lambda\to\infty}\exp\left(-i \alpha \sum_k\langle{\psi^5_k}|\gamma_5 |\psi^5_k\rangle \exp(-{\lambda^5_k}^2/\Lambda^2)\right)  \ ,\nn
\ee
where $\Lambda$ is the width of the regularization. Subsequently, he reformulates  the regulator to show that the sum equals the topological index $\nu$ of the gauge field configuration.
We follow a different path and consider the quantity
\be
  \sum_k\delta(\lambda^5-\lambda^5_k)\langle{\psi^5_k}|\gamma_5 |\psi^5_k\rangle
\ee 
which allows us to turn the sum in the exponent of $J$ into an integral. To understand the integrand in the exponent, we introduce the ensemble averaged distribution of the chiralities over the spectrum of $D_5$  
\be
\rho_{5,\chi}(\lambda^5)  \equiv \left\langle \sum_k\delta(\lambda^5-\lambda^5_k)\langle{\psi^5_k}|\gamma_5 |\psi^5_k\rangle\right\rangle \ .
\ee
We have the relation
\be
\rho_{5,\chi}(\lambda^5) 
& = & \frac{1}{\pi}{\rm Im}\left[\Sigma(m,z=-\lambda^5)\right] \ , \nn
\ee
where the Green function of $D_W$ is
\be
\Sigma(m,z)\equiv\left\langle \tr\frac{1}{D_W+m+z\gamma_5}\right\rangle = \left\langle \tr\frac{\gamma_5}{D_5+z}\right\rangle .
\ee
Hence we may compute $\rho_{5,\chi}$ from the EFT by the supersymmetry technique as outlined in Appendix \ref{app:SUSY}. The outcome is most interesting: The resulting $\rho_{5,\chi}(\lambda^5)$ has a $1/\lambda^5$ tail for $\lambda^5\gg m,a$. Hence, the Riemannian way of integrating it would produce a logarithmic divergence instead of reproducing the topological index $\nu$.
One needs to understand the integral over $\rho_{5,\chi}(\lambda^5)$ like a principal value integral, in particular we need to integrate $\int_0^\infty\rho_{5,\chi}^{\rm (s)}(\lambda^5)d\lambda^5=\nu$ with 
\be
\rho_{5,\chi}^{\rm (s)}(\lambda^5)\equiv\rho_{5,\chi}(\lambda^5)+\rho_{5,\chi}(-\lambda^5) \ .
\ee
 The integrand is then completely dominated by the topological peak, as illustrated in figure 1, where we  also plot the level density $\rho_5(\lambda^5)$ of $D_5=\gamma_5(D_W+m)$~\cite{DSV}.
We stress that this is fully consistent and the natural extension of Fujikawa's result.


\end{document}